\documentclass{iaus}
\usepackage{graphicx}

\title[Transit observation of XO-1b and TrES-1] 
{Transit observation at the observatory in Gro{\ss}schwabhausen: XO-1b and
TrES-1}

\author[M. Va{\v n}ko]   
{M. Va{\v n}ko$^1$, St. Raetz$^1$, M. Mugrauer$^1$, T.O.B. Schmidt$^1$, 
T. Roell$^1$, T.~Eisenbeiss$^1$, M. Hohle$^1$, A. Seifahrt$^{1,2}$, A.
Koeltzsch$^1$, Ch. Broeg$^{1,3}$, J. Koppenhoefer$^4$
 \and R. Neuh{\"a}user$^1$}

\affiliation{$^1$Astrophysikalisches Institute und Universit\"ats-Sternwarte,\\ 
Schillerg\"a{\ss}chen 2-3, 07745 Jena, Germany \\ email: {\tt vanko@astro.uni-jena.de} \\[\affilskip]
$^2$Institut f\"ur Astrophysik, Georg-August-Universit\"at, \\ 
Friedrich-Hund-Platz 1, 37077, G\"ottingen, Germany \\[\affilskip] 
$^3$Space Research and Planetary Sciences, Physikalishes Institute, University of Bern,\\ 
Sidlerstra{\ss}e 5, 3012 Bern, Switzerland \\[\affilskip]
$^4$Max-Planck Institute of Extraterrestrial Physics,\\
Giessenbachstra{\ss}e, 85748 Garching, Germany \\[\affilskip]} 

\pubyear{2008}
\volume{253}  
\pagerange{119--126}
\jname{Title of your IAU Symposium}
\editors{A.C. Editor, B.D. Editor \& C.E. Editor, eds.}
\begin{document}

\maketitle

\begin{abstract}
We report on observations of transit events of the transiting planets
XO-1b and TrES-1 with the AIU Jena telescope in Gro{\ss}schwabhausen. Based on
our ($IR$) photometry (in March 2007) and available transit timings
(SuperWASP, XO and TLC-project-data) we improved the orbital period of 
XO-1b ($P$ = 3.941497$\pm$0.000006 ) and TrES-1 ($P$ = 3.0300737$\pm$0.000006), 
respectively. The new ephemeris for the both systems are presented.   

\keywords{binaries: eclipsing, planetary systems, stars: individual (GSC
02652-01324,\\ GSC~02041~-~01657), techniques: photometric}
\end{abstract}

\firstsection 
\section{Introduction}

The transiting planets are giving us an opportunity to obtain important 
information about both the planet and the star. With precise measurements
of transit events it is possible to infer the relative size of the star and
planet, the orbital inclination and the stellar limb-darkening function.
Having spectroscopic measurements of the time-variable Doppler shift of the
star and an estimate of the stellar mass we can determine planetary mass and
the stellar radius.\\
In this paper we present observations of two known transiting planets, XO-1b and
TrES-~1. The observations were carried out with the AIU Jena telescope at the
observatory in Gro{\ss}schwabhausen on March 2007. We have used the 25cm
Cassegrain telescope installed at the tube of the 0.9m telescope. The $R$
and $I$-band images were obtained with the Cassegrain-Telescope-Kamera (CTK)
a CCD-camera with 37.7'x37.7' field of view and 1024x1024 pixels.
(\cite[Mugrauer et al. (2008)]{Mugrauer_etal08}, in preperation).\\
For both systems, XO-1b and TrES-1, we have observed one transit event in
$I$-band and $R$-band, respectively. Obtained data were reducted by standard
IRAF procedures. For data analysis we have used IRAF task {\it chphot},
written by Ch. Broeg and based on the standard IRAF routine {\it phot}
(\cite[Broeg et al. 2005]{Broeg_etal05}). The IRAF task {\it chphot}
includes an algorithm which uses as many stars as possible and calculates 
one artificial comparison star (hereafter CS). This process is based on 
weighted averages of CSs, where an algorithm selects the most stable stars 
and computes the one artificial CS with the best S/N ration.
Finally, we have corrected data on
systematic effects using the "Sys-Rem" detrending algorithm which is
proposed by \cite[Tamuz et al. (2005)]{Tamuz_etal05} and implemented by
Johannes Koppenhoefer. The main aim of our investigation is to improve
orbital periods of these transiting planets as well as to discuss possible 
changes of orbital periods based on their O-C diagrams. 
For this purpose we collected all available transit times for each transiting system
using SuperWASP, XO and TLC-project data. 



\section{XO-1b}

We have observed the transit of the exoplanet XO-1b on March 11th 2007.
According to the ephemeris provided by \cite[McCullough et al.
(2006)]{McCullough_etal06}:

\begin{equation}
T_{\mathrm{c}}(E) = (2453808.9170 + E\cdot 3.941534)d,
\end{equation}
this transit corresponds to epoch 92. During observation we have performed 
161 $I$-band exposures. Our mean photometric accuracy is 0.008 mag. Due to
the variable observational conditions, the photometric accuracy at the
beginning of the night was a bit worse then at the end.  	
For data analysis we have used aperture photometry on all available images. 
To get, as good as possible transiting light curve, we have used only
the most stable CSs in the field. After the first run of the
artificial-comparison-star-algorithm we have rejected 168 CSs. 
\begin{figure}[h]
  \centering
  \includegraphics[width=1.025\textwidth]{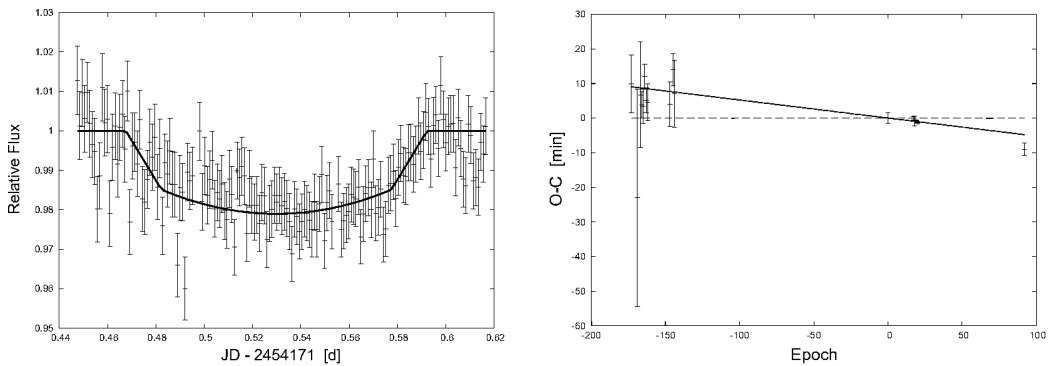}
  \caption{({\it left plot}) Relative $I$-band photometry of XO-1b. 
({\it right plot}) Transit timing residuals for XO-1b. The dashed line shows
the ephemeris given by \cite[McCullough et al. (2006)]{McCullough_etal06}.
The best fit (dotted line) is representing by the updated ephemeris.}
\end{figure}
After repetition of the algorithm we have rejected other faint stars with low
S/N and stars suspected of variability. With the remaining 28 most stable
stars we calculated the artificial CS. In the resulting light curves we have used Sys-Rem. 
The algorithm works without any prior knowledge of the effects. The number of effects that
should be removed from the light curves is selectable and can be set as a
parameter. Using Sys-Rem with two effects the transit itself is
disappearing from the light curve. Thus we have used only one effect.      

To determine the time of the center transit we have used the fit based on
the system parameters of \cite[Holman et al. (2006)]{Holman_etal06}. With
the help of the {$\chi^2$}-test, we have determined the time of the
midtransit as follows:

\begin{equation}
T_{\mathrm{c}}(HJD) = (2454171.53188 \pm 0.00130)d.
\end{equation}

The final time series is plotted in Fig.~1 (left plot). Except of our transit,
observed at the observatory in Gro{\ss}schwabhausen, we found 16 other 
transit times for XO-1b in the literature. Together with our transit, all 17
points allow us to study the (O-C) diagram and discuss possible changes of
period. We have used the ephemeris of \cite[McCullough et al. (2006)]{McCullough_etal06}
to compute "Observed minus Calculated" (O-C) residuals for all 17 transit
times. Figure~1 (right plot) shows the differences between the oberved and
predicted times of midtransit as a function of epoch. The dashed line
represents the ephemeris given by \cite[McCullough et al. (2006)]{McCullough_etal06}. 
We found a negative trend in this (O-C) diagram.
For an exact determination of the orbital period we set the transit time
with the smallest uncertainties as $T_{\rm 0}$. In this case we have used
the transit with epoch 20, according to the ephemeris of \cite[McCullough et
al. (2006)]{McCullough_etal06}, observed in the TLC-Project. Hence, the transit
time for the epoch 0 is $T_{\mathrm{c}}$(HJD) = (2453887.74679 $\pm$ 0.00015)d.  
Therefore we plotted the midtransit times over the epoch and did a linear fit
with fixed $T_{\rm 0}$. We got the best {$\chi^2$} with an orbital period of
$P$ = (3.941497$\pm$0.000006) days. Within the error bars almost 70\% of the
points are consistent with our calculated period. The remaining measurements
are less than 1-$\sigma$ from the new "zero" line (solid line in Fig. 1 - right plot). 
The resulting ephemeris which is in a good agreement with our observation is:

\begin{equation}
T_{\mathrm{c}}(HJD) = (2453887.74679 + E\cdot 3.941497)d.
\end{equation}

\section{TrES-1}

The other observation at the observatory in Gro{\ss}schwabhausen
was performed on March 15th 2007. We observed one transit of TrES-1 with 
the same 24.5cm Cassegrain telescope. The light curve consists of 88 $R$-band
images with 60s exposures. Unfortunately, our observation was aborted too early 
and the last part of the light curve is missing. This transit corresponds 
to epoch 326 of the ephemeris given by \cite[Winn et al.(2007)]{Winn_etal07}:

\begin{equation}
T_{\mathrm{c}}(E) = (2453186.80603 + E\cdot 3.0300737)d.
\end{equation}
 
In this case our mean photometric accuracy is 0.009 mag. The data reduction
and analysis was carried out in the same way as in the case of XO-1b. For
calculation of the artificial CS we have used 34 most stable stars with a good
S/N. As we did not have light from out of transit which allow us to find the systematic
effect, the Sys-Rem was not used. The determination of the transit center
was done in the same way as in the case of XO-1b. After normalization we did
a fit of the light curve using the system parameters by \cite[Winn et al.
(2007)]{Winn_etal07}. Using the theoretical light curve and the
{$\chi^2$}-test, it was possible to estimate the center of the transit even
without the egress. The minimal value of {$\chi^2$} corresponds with
the following midtransit time:

\begin{equation}
T_{\mathrm{c}}(HJD) = (2454174.60958 \pm 0.00150)d.
\end{equation}
 
The resulting light curve is shown in the Fig. 3 (left plot).
For TrES-1 we found 15 midtransit times in the literature. The transit of
epoch 40.5 (forced $e$ = 0) according to the ephemeris given by \cite[Winn
et al. (2007)]{Winn_etal07} is even a secondary transit observed by
\cite[Charbonneau et al. (2005)]{Charbonneau_etal05}. With all 16 available
transit times we determined the transit and secondary-eclipse timing
residuals for TrES-1. The calculated times (using the ephemeris of
\cite[Winn et al. (2007)]{Winn_etal07}) have been subtracted from the
observed times. The resulting (O-C) diagram is shown in the Fig.~2 (right
plot). The point lie on the horizontal line. It means that data are
consistent with a constant period which confirmes the ephemeris given by
\cite[Winn et al. (2007)]{Winn_etal07}.    
\begin{figure}[h]
  \centering
  \includegraphics[width=1.025\textwidth]{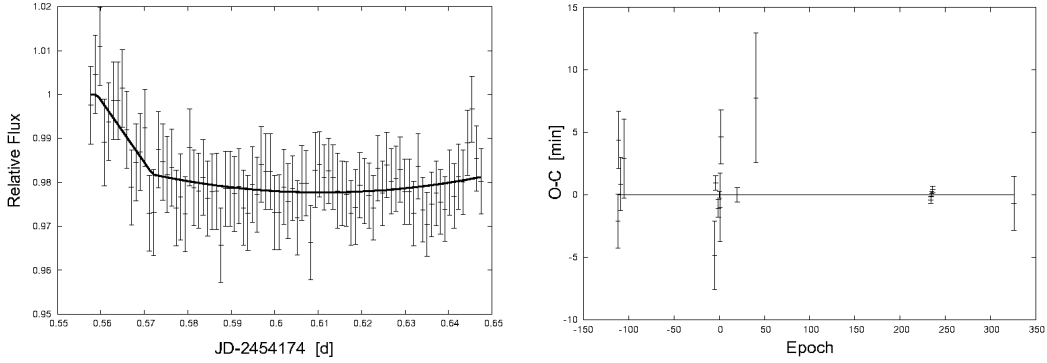}
  \caption{({\it left plot}) Relative $R$-band photometry of TrES-1.
({\it right plot}) Transit timing residuals for TrES-1. The dashed line
shows the ephemeris given by \cite[Winn et al. (2007)]{McCullough_etal07}.
The data points are consistent with the constant period.}
\end{figure}

\section{Conclusions and Discussion}

During March 2007 we have observed transits of known transiting planets
XO-1b and TrES-1 at the university observatory in Gro{\ss}schwabhausen. 
Using the theoretical light curve and {$\chi^2$}-test we determined the time
of the midtransit for XO-1b ($T_{\mathrm{c}}$(HJD) = (2454171.53188 $\pm$
0.00130)d) and TrES-1 ($T_{\mathrm{c}}$(HJD) = (2454174.60958 $\pm$
0.00150)d). Using our photometry and other transit timing from literature 
we improved the orbital period XO-1b ($P$ = 3.941497$\pm$0.000006 ) 
and TrES-1 ($P$ = 3.0300737$\pm$0.000006), respectively. The same transit timing
we have used for the construction of the (O-C) diagrams of both transiting
systems. In the (O-C) diagram of XO-1b we found a negative trend.
In the Fig. 1 (right plot) the dashed line shows ephemeris
given by \cite[McCullough et al. (2006)]{McCullough_etal06} and the
best-fitting solid line is representing the updated ephmemeris given in the
follow equation: $T_{\mathrm{c}}$(HJD) = (2453887.74679 + E$\cdot$ 3.941497)d.
For TrES-1 we have found 15 transit times (with our data there are 16
transits) and we constructed a (O-C) diagram (see Fig. 2 - right plot). As we
can see, the points lie on a horizontal line. This means confirmation of
the ephemeris given by Winn et al. (2007).

Therefore, we try to develop a method for the accuracy improvement of our
transit times, we would like to continue in our observations of transiting
exoplanets at the observatory in Gro{\ss}schwabhausen. In addition, 
in the future, we plan to use the 90cm reflector telescope and improve our
observations.

\end{document}